\documentclass[%
 reprint,
 amsmath,
 amssymb,
 aps,
 nofootinbib
]{revtex4-2}

\usepackage{physics}
\usepackage{graphicx}
\usepackage{dcolumn}
\usepackage{bm}
\usepackage[parfill]{parskip}
\usepackage[usenames,dvipsnames]{xcolor}
\usepackage{hyperref}
\usepackage{cleveref}
\usepackage{nameref}
\hypersetup{
    pdfencoding=unicode,
	colorlinks=true,
	urlcolor=Maroon,
	linkcolor=RoyalBlue,
	citecolor=Maroon,
	pdftitle={Recursive Landau Analysis},
	pdfauthor={Simon Caron-Huot, Miguel Correia, Mathieu Giroux},
	pdfdisplaydoctitle=true,
	pdfstartview=FitH,
	linktocpage=true
}
\usepackage{amsmath}
\usepackage{dsfont}
\usepackage{amsfonts}
\usepackage{amssymb}
\usepackage{mathtools}
\usepackage{microtype}
\usepackage{bm}
\usepackage[utf8]{inputenc}
\usepackage{blkarray}
\usepackage{bigstrut}
\usepackage{nccmath}
\usepackage{float}
\usepackage{braket}
\usepackage{dsfont}
\usepackage[export]{adjustbox}
\usepackage[english]{babel}
\usepackage{stmaryrd}
\usepackage{orcidlink}
\usepackage{tikz}
\usepackage{bbm}
\usepackage{blindtext}
\usepackage{fontawesome}
\usepackage[]{notes2bib}
\usepackage{enumitem}
  
\usepackage{titlesec}
\titlespacing*{\section}{0.4pt}{0.2\baselineskip}{0.3\baselineskip}
\titleformat{\section}
 {\centering\bfseries\scshape}{\thesection}{1em}{}
\bibnotesetup{
note-name = ,
use-sort-key = false
}
\usepackage[caption=false]{subfig}
\usetikzlibrary{decorations.markings,decorations.pathmorphing}
\renewcommand{\thesection}{\arabic{section}}
\usepackage[skip=5pt,font=small,labelfont=bf,
   justification=Justified,
   format=plain]{caption}

\usepackage{ntheorem}
\usepackage[mathscr]{euscript}
\usepackage[framemethod=tikz,footnoteinside=false,hidealllines=true,topline=true,bottomline=true,linecolor=black!10!white,linewidth=1pt,innerleftmargin=0em,innerrightmargin=0em,frametitlerule=true,ntheorem]{mdframed}
\newcommand{\myparagraph}[1]{\paragraph*{\text{\textsc{#1}}}}

\usetikzlibrary{decorations.pathmorphing,decorations.pathreplacing,calligraphy,calc,snakes,cd,external,arrows.meta}
\tikzset{
    vector/.style={
        decoration={snake, aspect=0.75, mirror, segment length=2mm},
        decorate
    },
	photon/.style={decorate, decoration={snake, amplitude=1pt, segment length=6pt}
	}
}

\newcommand{\red}[1]{{\textcolor{black}{#1}}}
\newcommand{\redd}[1]{{\textcolor{black}{#1}}}

\def\D{\mathrm{D}}
\def\d{\mathrm{d}}

\DeclareFontFamily{U}{mathx}{}
\DeclareFontShape{U}{mathx}{m}{n}{<-> mathx10}{}
\DeclareSymbolFont{mathx}{U}{mathx}{m}{n}
\DeclareMathAccent{\widehat}{0}{mathx}{"70}
\DeclareMathAccent{\widecheck}{0}{mathx}{"71}
 
\DeclareMathOperator*{\sumint}{%
\mathchoice%
  {\ooalign{$\displaystyle\sum$\cr\hidewidth$\displaystyle\int$\hidewidth\cr}}
  {\ooalign{\raisebox{.14\height}{\scalebox{.7}{$\textstyle\sum$}}\cr\hidewidth$\textstyle\int$\hidewidth\cr}}
  {\ooalign{\raisebox{.2\height}{\scalebox{.6}{$\scriptstyle\sum$}}\cr$\scriptstyle\int$\cr}}
  {\ooalign{\raisebox{.2\height}{\scalebox{.6}{$\scriptstyle\sum$}}\cr$\scriptstyle\int$\cr}}
}

\makeatletter
\newenvironment{sqcases}{
  \matrix@check\sqcases\env@sqcases
}{
  \endarray\right.
}
\def\env@sqcases{
  \let\@ifnextchar\new@ifnextchar
  \left\lbrack
  \def\arraystretch{1.2}
  \array{@{}l@{\quad}l@{}}
}
\makeatother
\usepackage{pgfplots}
\pgfplotsset{compat=1.18}
\usepgfplotslibrary{fillbetween}
\usetikzlibrary{patterns}

\tikzset{ma/.style={decoration={markings,mark=at position 0.5 with {\arrow[scale=0.7]{>}}},postaction={decorate}}}
\tikzset{ma2/.style={decoration={markings,mark=at position 0.1 with {\arrow[scale=0.7,Maroon!70!black]{>}}},postaction={decorate}}}
\tikzset{ma3/.style={decoration={markings,mark=at position 0.7 with {\arrow[scale=0.7,RoyalBlue!80!black]{>}}},postaction={decorate}}}
\tikzset{mar/.style={decoration={markings,mark=at position 0.5 with {\arrowreversed[scale=0.7]{>}}},postaction={decorate}}}
\begin{document}
\allowdisplaybreaks
\title{Recursive Landau Analysis}
\author{Simon Caron-Huot$\,^{\orcidlink{0000-0002-7005-9652}}$,
Miguel Correia$\,^{\orcidlink{0000-0001-6636-9570}}$,
Mathieu Giroux$\,^{\orcidlink{0000-0002-2672-634X}}$
}

\affiliation{Department of Physics, McGill University, 3600 Rue University, Montr\'eal, H3A 2T8, QC Canada\\
\textnormal{\href{mailto:schuot@physics.mcgill.ca}{\texttt{schuot@physics.mcgill.ca}},
\href{mailto:miguel.ribeirocorreia@mcgill.ca}{\texttt{miguel.ribeirocorreia@mcgill.ca}},
\href{mailto:mathieu.giroux2@mail.mcgill.ca}{\texttt{mathieu.giroux2@mail.mcgill.ca}}}
}

\begin{abstract}
We propose a recursive method that makes use of the basic principle of unitarity to calculate the Landau singularities of $n$-point scattering amplitudes directly in kinematic space. For a vast class of Feynman diagrams, the method enables rapid analytic computation of Landau singularities beyond current state-of-the-art technology. This includes new predictions relevant for two- and higher-loop processes in the Standard Model involving both massive quarks and electroweak particles.
\end{abstract}

\maketitle

\allowdisplaybreaks
\raggedbottom
\section{\label{sec:intro}Introduction}
Scattering amplitudes are fascinating objects \cite{Dixon:2011xs} that characterize the interactions of particles and can be used to determine cross-sections and other useful physical observables.
They exhibit a rich analytic structure as a function of the momenta of the scattered particles.

In weakly coupled quantum field theories, such as the Standard Model at high energies, the Feynman diagram expansion offers a systematic way to organize calculations of amplitudes and to study their analytic structure. \red{However}
, the computation of Feynman integrals remains a difficult task that often becomes intractable at high orders, particularly when massive particles are involved (see \cite{Bourjaily:2022bwx} for a recent overview).

Knowing beforehand the singularity structure of a Feynman integral \red{can} greatly help in its computation and may even suffice to ``bootstrap'' the result when combined with additional physical constraints and a good control over the function space (e.g., polylogaritms and their elliptic analogues \cite{Weinzierl:2022eaz}). This principle has proven to be especially rewarding for planar $\mathcal{N}=4$ super Yang--Mills amplitudes 
\cite{Drummond:2018caf,Dixon:2020bbt,Caron-Huot:2016owq,Caron-Huot:2019vjl,Dixon:2020cnr,Dixon:2021tdw,Dixon:2023kop} \red{and it is also starting to be used in more realistic theories \cite{Chicherin:2020oor,Wilhelm:2022wow,Morales:2022csr,Cao:2023tpx,He:2023umf,Jiang:2024eaj,Giroux:2024yxu,Abreu:2024flk, Hannesdottir:2024hke}.}

The singularity locus of a Feynman integral can be studied without having to first compute \red{the integral explicitly}, as Landau proposed in his seminal paper \cite{Landau:1960jol}. The \emph{Landau equations} \cite[Eq.~(1)]{Fevola:2023fzn} are a set of necessary algebraic conditions for the existence of a singularity of the Feynman integral. Unsurprisingly, their complexity also increases with the number of loops, external particles, and mass scales. The current state-of-the-art makes use of modern algebraic geometry techniques to compute two-loop examples relevant for Standard Model processes \cite{Mizera:2021icv,Fevola:2023fzn,Fevola:2023kaw}
(see \cite{Flieger:2022xyq,Jiang:2024eaj,Bourjaily:2022vti,Lippstreu:2022bib,Lippstreu:2023oio,Klausen:2023gui,Dlapa:2023cvx,Helmer:2024wax,Correia:2021etg,Mizera:2022dko,Berghoff:2022mqu,Hannesdottir:2022xki}).

The Landau singularities of Feynman integrals are purely kinematical and are believed to exist beyond the weak-coupling approximation that underlies the Feynman diagrammatic expansion. For example, it was noted in the 1960s by Mandelstam and others \cite{Mandelstam:1959bc,Gribov:1962ft,kim} (see \cite{Correia:2020xtr, Correia:2021etg, Correia:2022dcu, Tourkine:2023xtu} for recent accounts) that general principles such as unitarity, \red{crossing symmetry and causality \cite{Iagolnitzer:1969sk,Iagolnitzer:1994xv}} require non-perturbative amplitudes to be singular at the same locations as the Landau singularities found for some Feynman diagrams. 

Here, we bring together valuable insights from both \red{perturbative and nonperturbative} approaches. We propose a recursive method for the calculation of Landau singularities exploiting the general principle of unitarity.
\section{\label{sec:algo}Recursion via unitarity}
A Feynman \red{integral associated with a} graph $G$ with $n$ external legs is a function on the \emph{kinematic space} consisting of Lorentz-invariant combinations of external momenta $X_G=\{p_i{\cdot}p_j\}_{i,j=1}^{n-1}$ and internal masses. Its codimension-one singularities \red{can be} characterized by a list ${\cal L}(G)\red{=\{ {\cal L}(G)_i\}}$ of \red{irreducible} polynomials in these variables
\begin{align}
 {\cal L}(G)_i = 0\,.
\end{align}
Our goal is to compute these polynomials recursively in terms of those of subgraphs. The product $\prod_i {\cal L}(G)_i$ is called the (principal) \emph{Landau discriminant} in \cite{Mizera:2021icv,Fevola:2023fzn,Fevola:2023kaw}. Before discussing complex singularities, we start by revisiting the relation between singularities and \red{graph} cuts for real momenta.
\vspace{0.1cm}
\myparagraph{Unitarity and thresholds.}

\emph{Unitarity} is the basic quantum mechanical requirement that probabilities sum up to unity. In terms of the S-matrix, which evolves states from the far past to the far future, it reads $SS^\dagger=\mathbbm{1}$. By the usual separation between free and interacting parts ($S{=}\mathbbm{1}{+}iT$), it implies that $2\mathrm{Im} \, T = T T^\dagger$.

\emph{Scattering amplitudes} are the matrix elements of $T$: $\mathscr{M}_{n_A \to n_B} = \langle n_B | T | n_A \rangle$, where $|n_A \rangle$ and $|n_B \rangle$ denote $n_A$- and $n_B$-particle momentum eigenstates. In terms of matrix elements, unitarity yields an infinite system of coupled integral equations (see e.g., \cite[Ch.~2]{Hannesdottir:2022bmo})
\begin{equation}
     2\mathrm{Im} \, \mathscr{M}_{n_A \to n_B} = \sumint_X \mathscr{M}_{n_A \to X}~\mathscr{M}^*_{X \to n_B}\,,
     \label{eq:unitarity}
\end{equation}
where the integral is over the on-shell phase space of the intermediate $X$-particle state \red{and $\mathscr{M}^*$ denotes the complex conjugated amplitude}. In perturbation theory, \eqref{eq:unitarity} leads to the well-known \emph{Cutkosky equation} \cite{cutkosky1961anomalous}, which relates the imaginary part of a given Feynman diagram to the sum over all cuts that divide it into two or more disconnected subgraphs; each cut particle being \emph{on-shell}.

Here we will consider the case where a two-particle cut separates a graph $AB$ into two subgraphs $A$ and $B$. The invariants on each side of the cut are
\begin{equation}
      X_\xi=\{q_i\cdot  q_j\mid q_\bullet\in \{k\}\cup P_\xi\} \quad (\xi= A,B)\,,
\end{equation}
where $P_A=\{p_i\}_{i=1}^{a}$ and $P_B=\{p_i\}_{i=a+1}^{n}$. Below, we also adopt $p_I$ to denote the sum $p_I\equiv\sum_{i\in I} p_i$ (where $I$ is a multi-index) and $s_{I}\equiv p_I^2$ for Mandelstam invariants. 
We also ignore numerators and spin, as they do not affect the locations of possible Landau singularities.

Following Mandelstam \cite{Mandelstam} or the more general Baikov representation \cite{Baikov:1996iu,Mastrolia:2018uzb}, the integrals \red{in \eqref{eq:unitarity}} can be written in terms of these invariants (assuming initially that positive energy-flow along the cut is possible, see \cite[Eq.~(2.29)]{Hannesdottir:2022bmo}). \red{In these variables the Cutkosky relation for a two-particle cut takes the form}

\begin{equation}
\begin{split}
&\adjustbox{valign=c,scale={0.85}{0.85}}
{
\includegraphics[scale=1]{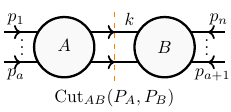}
}
\hspace{-0.1cm}
= C\int_\Gamma \d \mu {A(X_A) \, B(X_B) \over \det \mathcal{G}(Q)^{\frac{n+1-\D}{2}}}\,,
    \\&
    \hspace{0.1cm}
    \d \mu=\prod_{Q_i\in Q}\d (k\cdot Q_i) \delta[k^2{-}m_k^2] \, \delta[(k{-}p_{1\ldots a})^2{-}m_{k'}^2]\,,
    \\&
    \hspace{0.1cm}
    \phantom{d}C=\det \mathcal{G}(P_A{\cup}P_B')^{\frac{n-\D}{2}}/\sqrt{\pi}\,, ~ Q=\{k\}{\cup}P_A{\cup}P'_B
    \end{split} \label{eq:cut}
\end{equation}
where $P_B'=P_B\setminus \{p_n\}$ and $\mathcal{G}(P)$ denotes the \emph{Gram matrix} of dot products between vectors in $P$.  The integration domain is
\begin{equation}\label{eq:contourDef}
    \Gamma=\Big\{k\cdot Q_i \,\Big|\,\frac{\det \mathcal{G}(Q)}{\det \mathcal{G}(P_A{\cup}P_B')}>0, Q_i\in Q\Big\}\,.
\end{equation}
The delta functions in \eqref{eq:cut} can be used to fix two scalar products in $X_A$ (one being $k^2$). Since there are $n_A{=}a$ initial particles and $n_B{=}(n{-}a)$ final particles, the final integration is over $(n_A {+} n_B {-} 2)$ variables.

The locations in kinematic space where a cut such as \eqref{eq:cut} starts contributing to \eqref{eq:unitarity} are called \emph{thresholds}. At such locations, the amplitude \red{is not} analytic (it has a pole or branch point) and \red{one says} that it is \emph{singular}.

Landau's analysis shows that all singularities of a diagram are associated with processes in which a subset of its propagators become on-shell. \red{In this work} we will focus on the leading singularities of a graph $G$: those in which all propagators are on-shell.
\red{We will focus on graphs that contain a two-particle cut described by 
\eqref{eq:cut}-\eqref{eq:contourDef}, which define an analytic function when continued away from real kinematics.} 
Necessary conditions for it to become singular will be phrased algebraically without reference to the positivity of energies or the reality of Mandelstam invariants.

\vspace{0.1cm}
\myparagraph{Necessary conditions for singularities.}


\red{As described in \cite{Eden:1966dnq}, singularities of integral representations can be generated by two mechanisms: 1. Two or more singularities of the integrand coincide and trap the contour of integration, preventing its deformation.  2. A singularity of the integrand moves to the boundary of integration.
These are known as \emph{pinch} and \emph{end-point} singularities.
For the integral \eqref{eq:cut}, we can say that these two possibilities involve singularities of $A$ and/or $B$ colliding among themselves or with the boundary $\partial \Gamma$.
Here, since our integration boundary $\partial\Gamma$ is not fixed, we must also consider the possibility that $\partial\Gamma$ collides with itself, i.e. becomes degenerate.
This happens, for example, at the thresholds mentioned above where the phase space $\Gamma$ reduces to a single point.
}

A fourth, trivial, possibility is that the pre-factor $C$ in \eqref{eq:cut} becomes singular.
The proposed necessary conditions for the cut $A B$ 
to develop a singularity are then:
\red{
\begin{enumerate}[nosep]
    \item[(a.1)] Singularities of $A$ and/or $B$ collide with each other. 
    \item[(a.2)] A singularity of $A$ or $B$ collides with $\partial \Gamma$.
    \item[(a.3)] The integration boundary $\partial\Gamma$ collides with itself. 
    \item[(b)] The pre-factor $C$ becomes singular.
\end{enumerate}
Visually, conditions (a.1-3) look like}
\begin{equation} \tikzset{every picture/.style={line width=0.75pt}}
\begin{tikzpicture}[x=0.75pt,y=0.75pt,yscale=-1,xscale=1]
\draw[->]    (107,111) -- (289.42,111) node[right,xshift=0.5cm]{$(\xi=A,B)$};
\draw    (244,61) -- (244,79.42) -- (280.42,79.42) ;
\begin{scope}
    
\end{scope}
\draw [color=Maroon  ,draw opacity=1,thick]   (148,111) -- (253,111) ;
\draw [shift={(253,111)}, rotate = 359.68] [color=Maroon  ,draw opacity=1 ][fill=Maroon  ,fill opacity=1 ][line width=0.75]      (0, 0) circle [x radius= 1.34, y radius= 1.34]   ;
\draw [shift={(148,111)}, rotate = 359.68] [color=Maroon  ,draw opacity=1 ][fill=Maroon  ,fill opacity=1 ][line width=0.75]      (0, 0) circle [x radius= 1.34, y radius= 1.34]   ;
\begin{scope}[yshift=-0.53cm,xshift=-0.53cm]
\draw [-{Triangle[scale=0.7]}] [color=Maroon]    (168,131) -- (220.21,130.71) ;
\draw [{Triangle[scale=0.7]}-] [color=Maroon]   (218.21,130.71) -- (272.42,130.42) ;
\draw  [color=RoyalBlue  ,draw opacity=1 ][fill=RoyalBlue  ,fill opacity=1 ] (156,105.71) .. controls (156,104.76) and (156.76,104) .. (157.71,104) .. controls (158.65,104) and (159.42,104.76) .. (159.42,105.71) .. controls (159.42,106.65) and (158.65,107.42) .. (157.71,107.42) .. controls (156.76,107.42) and (156,106.65) .. (156,105.71) -- cycle ;
\draw  [color=RoyalBlue  ,draw opacity=1 ][fill=RoyalBlue  ,fill opacity=1 ] (249,146.71) .. controls (249,145.76) and (249.76,145) .. (250.71,145) .. controls (251.65,145) and (252.42,145.76) .. (252.42,146.71) .. controls (252.42,147.65) and (251.65,148.42) .. (250.71,148.42) .. controls (249.76,148.42) and (249,147.65) .. (249,146.71) -- cycle ;
\draw[{Triangle[scale=0.7]}-] [color=gray!50  ,draw opacity=1 ]   (277.5,134.5) .. controls (295.5,146.5) and (291.87,153.58) .. (256.71,147.71) ;
\draw  [color=RoyalBlue ,draw opacity=1 ][fill=RoyalBlue ,fill opacity=1 ] (189,149.21) .. controls (189,148.26) and (189.76,147.5) .. (190.71,147.5) .. controls (191.65,147.5) and (192.42,148.26) .. (192.42,149.21) .. controls (192.42,150.15) and (191.65,150.92) .. (190.71,150.92) .. controls (189.76,150.92) and (189,150.15) .. (189,149.21) -- cycle ;
\draw[{Triangle[scale=0.7]}-] [color=gray!50  ,draw opacity=1 ]  (188.33,128) .. controls (195.33,114) and (144.29,128.29) .. (157,112.5);
\draw[-{Triangle[scale=0.7]}] [color=gray!50 ,draw opacity=1 ]   (185.33,149) .. controls (169.33,150) and (182.33,142) .. (186.33,136) ;
\end{scope}

\draw (247,62.4) node [anchor=north west][inner sep=0.75pt]    {$k{\cdot}P_\xi$};
\draw (112,65.4) node [anchor=north west][inner sep=0.75pt]  [color=RoyalBlue  ,opacity=1 ]  {$\mathscr{L}(\xi)=0$};
\draw (243,96.4) node [anchor=north west][inner sep=0.75pt]  [color=Maroon  ,opacity=1 ]  {$\partial\Gamma$};
\draw (270,122) node [anchor=north west][inner sep=0.75pt]   [align=left, color=gray!50] {(a.2)};
\draw (160.21,80.71) node [anchor=north west][inner sep=0.75pt]   [align=left, color=gray!50] {(a.1)};
\draw (190.5,115.92) node [anchor=north west][inner sep=0.75pt]   [align=left, color=Maroon] {(a.3)};
\end{tikzpicture}
\end{equation}
In the multi-variable integration space,
any one of the (a) conditions can be applied to each individual integration variable and combinations can occur.

Crucially, since the boundary $\partial\Gamma$ is determined by the polynomial $\det \mathcal{G}=0$, the condition (a.3) can be phrased as $\partial_{k {\cdot} p_i} \det \mathcal{G} {=} 0$ for all remaining integration variables $\{k \cdot p_i\}$ that are not fixed already by (a.1) or (a.2) (otherwise there would remain some integration variable that can be freely deformed away to avoid the singularity).

The 
(a) conditions for the cut $AB$ to have a singularity can thus be written uniformly by picking a possibly empty subset $\mathscr{S}\subset \mathscr{L}(A)\cup \mathscr{L}(B)$ of the singularities of the left and right sub-amplitudes. 
Localizing to $\{\mathscr{S}_i=0\}$ allows to eliminate $\dim(\mathscr{S})$ variables from \eqref{eq:cut},  leaving a set $X_\mathscr{S}$ of independent variables in terms of which $\partial\Gamma$ is determined by
\begin{equation} \label{eq:Gtilde}
 0=\det \tilde{\mathcal{G}}(X_{\mathscr{S}})\equiv \det \mathcal{G}|_{\{\mathscr{S}_i=0\}}\,.
\end{equation}
To ensure that there are no directions along which the integration contour could be deformed, we impose condition (a.3) on the remaining variables:
\begin{equation}\label{eq:LAB}
   \mathscr{L}(AB)_{\mathscr{S}}=0 :
\, \begin{sqcases}
 \;\det \tilde{\mathcal{G}} = 0 \vspace{1mm} \\
    \displaystyle {\partial \det \tilde{\mathcal{G}} \over \partial (k \cdot p_i)} = 0
\end{sqcases} \text{ for }  k{\cdot}p_i \in X_{\mathscr{S}}\,.
\end{equation}
Since there is always one more equation than unknowns in \eqref{eq:LAB}, evaluating one on the support of the others yields a polynomial equation $\mathscr{L}(AB)_{\mathscr{S}}=0$ in kinematic space.\footnote{In \eqref{eq:LAB} we include situations where 
more constraints are imposed using (a.1) than there are variables in the integral \eqref{eq:cut}, in which case $\det\tilde{\cal G}=0$ is not imposed. However, since $A$ and $B$ do not share any variable in the Baikov representation, the resulting conditions can only involve the individual Landau locii of each subdiagram separately and all solutions we found in this way were redundant.
\red{Note also that the conditions (a.1) could potentially be used to eliminate more than $\dim(\mathscr{S})$ variables if the variety $\{\mathscr{S}_i=0\}$ is singular, however we did not encounter this possibility.}
}

To find all candidate leading singularities of a graph $G$ that contains a two-particle cut, it suffices to consider the union of all sets $\mathscr{S}$ of candidate leading singularities of the sub-amplitudes on that cut.
In addition, we must account for the last possibility (b), which is algebraically given by $\det \mathcal{G}(P_A{\cup}P_B') = 0$ and yields the ``second-type'' singularities of \cite{Eden:1966dnq}.

Formula \eqref{eq:LAB} is our main result. It forms the basis for a recursive method which finds all candidate leading singularities of any two-particle-reducible diagram. We now illustrate it in practice for the two-loop parachute diagram. Other checks and new predictions are presented in Sec.~\ref{sec:examples} (see Tabs.~\ref{fig:checked} and \ref{fig:predictions}).
\section{The parachute split open}\label{ex:parachute1}
\myparagraph{Leading singularities.} As a first example, we consider the parachute diagram in generic four-point kinematic, where $p_i^2=M_i^2$ for each $i=1,2,3,4$ and $p_{12}^2=p_{34}^2=s$.

More explicitly, we first look at the cut
\begin{equation}\label{eq:example1}
    \adjustbox{valign=c,scale={1}{1}}{
    \includegraphics[scale=1]{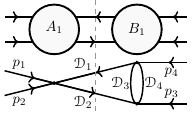}
    }=C_{\text{par}} \int_{\Gamma_1} {\d \mu_1 {A_1}\,  {B_1} \over (\det \mathcal{G}_1)^{\frac{4-\D}{2}}}\,,
\end{equation}
where
$\d \mu_1=\d(k_1{\cdot}p_{12})\d(k_1{\cdot}p_{3})\d(k_1^2) \delta[\mathscr{D}_1]\delta[\mathscr{D}_2]$,
\begin{equation}
    \mathscr{D}_1= (k_1-p_{12})^2-m_1^2\,,\qquad  \, \mathscr{D}_2=k_1^2-m_2^2 \,,
\end{equation} 
and 
\begin{equation}\label{eq:G2para}
     \mathcal{G}_1=\begin{bmatrix}
         p_{12}^2 & p_{12}\cdot k_1& p_{12}\cdot p_3\\
         p_{12}\cdot k_1 & k_1^2& k_1\cdot p_3\\
         p_{12}\cdot p_3 & k_1\cdot p_3 & p_3^2\\
     \end{bmatrix}\,.
\end{equation}
Above, ${A_1}$ is just a four-point vertex (e.g., ${A_1}=-i\lambda$ in $\phi^4$-theory) such that $\mathscr{L}(A_1)$ is the empty set, while ${B_1}$ is \red{the one-loop bubble integral 
with momentum $p_3 + k_1$ entering the bubble.} 

We show how localizing $\det\mathcal{G}_1=0$ on the two-particle cut \emph{and} the singular locus $\mathscr{L}(B_1)$ of $B_1$ where all propagators are cut gives the leading singularity of the parachute. This corresponds to condition (a.2) from Sec.~\ref{sec:algo}.

To find $\mathscr{L}(B_1)$, we exploit the recursive nature of \eqref{eq:cut} on the $B_1$ (bubble) blob, which gives
\begin{equation}\label{eq:example2}
    \adjustbox{valign=c,scale={1}{1}}{
    \includegraphics[scale=1]{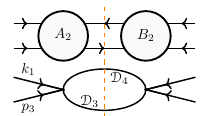}
    }=C_{\text{bub}} \int_{\Gamma_2} {\d \mu_2 {A_2}\,  {B_2} \over (\det \mathcal{G}_2)^{\frac{3-\D}{2}}}\,,
\end{equation}
where both ${A_2}$ and ${B_2}$ are (irrelevant) vertices,
\begin{equation}
    \mathscr{D}_3= (k_2+\Lambda)^2-m_3^2\,,\qquad \,\mathscr{D}_4=k_2^2-m_4^2\,,
\end{equation}
with $\Lambda{=}p_3{+}k_1,$ $\d \mu_2{=}\d(k_2{\cdot} \Lambda)\d(k_2^2) \delta[\mathscr{D}_3]\delta[\mathscr{D}_4]$  and 
\begin{equation}\label{eq:G1para}
     \mathcal{G}_2=\begin{bmatrix}
         \Lambda^2 & \Lambda\cdot k_2\\
         \Lambda\cdot k_2 & k_2^2&
     \end{bmatrix} \,.
\end{equation}
On the two-particle cut, \eqref{eq:example2} is singular when $\det \tilde{\mathcal{G}}_2=\det \mathcal{G}_2\vert_{\mathscr{D}_3{=}\mathscr{D}_4{=}0}=0$.
Hence, accounting for the pre-factor singularity $C_{\text{bub}}\propto(\Lambda^2)^{\frac{2-\D}{2}}$ (condition (b) above), the singular loci of the bubble are characterized by the irreducible polynomials
\begin{equation}\begin{aligned}\label{eq:condF1}
\mathscr{L}(B_1)&=\{
(k_1+p_3)^2-(m_3+m_4)^2,
\\&\!\!\! (k_1+p_3)^2-(m_3-m_4)^2, (k_1+p_3)^2\}\,.
\end{aligned}\end{equation}
Taking \eqref{eq:Gtilde} with $\mathscr{S}=\{\mathscr{L}(B_1)_1\}$ or $\{\mathscr{L}(B_1)_2\}$ (i.e., imposing one of \eqref{eq:condF1} on $\det \mathcal{G}_1\vert_{\mathscr{D}_1=\mathscr{D}_2=0}$ to eliminate $k_1{\cdot}p_3$)
gives $\det \tilde{\mathcal{G}}_1$. The result is free of $k_1$ so there is no derivative condition to impose. Thus, using $p_{12}{\cdot}p_3=\frac{1}{2}(M_4^2{-}M_3^2{-}s)$, $\mathscr{L}(A_1B_1)_{\mathscr{S}}=0$ reads
\begin{equation}\label{eq:detToSolve}
\hspace{-0.3cm}
\small
    \begin{vmatrix}
 s & \frac{m_2^2{-}m_1^2{+}s}{2} & \frac{M_4^2{-}M_3^2{-}s}{2} \\
 \frac{m_2^2{-}m_1^2{+}s}{2} & m_2^2 & \frac{(m_4{+}m_3)^2{-}m_2^2{-}M_3^2}{2} \\
 \frac{M_4^2{-}M_3^2{-}s}{2} & \frac{(m_4{+}m_3)^2{-}m_2^2{-}M_3^2}{2}& M_3^2
 \end{vmatrix}=0\,,
\end{equation}
where $|...|$ is a shorthand for the determinant. Expanding the determinant,
one finds a perfect match with the polynomial $\texttt{D[5]}$ in the \texttt{PLD.jl} database \cite{data1,Fevola:2023fzn}.

Notice that the fifth component is not the only leading singularity flagged by \texttt{PLD.jl}: there is also $\texttt{D}[2]$. Here, it \red{is captured by condition (b) above corresponding to}  $\mathscr{L}(B_1)_2=C_{\text{par}}\propto (\det[p_i\cdot p_j]_{i,j=1,2,3})^{\frac{3-\D}{2}}$ in \eqref{eq:example1}.

It was noted in \cite{Fevola:2023fzn,Fevola:2023kaw} that \texttt{PLD.jl} does not detect the singularity $\texttt{D}[3]$ identified by \texttt{HyperInt} \cite{Panzer:2014caa} (see also \cite[Eq.~(6.15)]{Berghoff:2022mqu}). Here, it is captured by starting the recursion using
the third entry of \eqref{eq:condF1} which represents the second-type singularity of the one-loop bubble integral.
In this case, $\mathscr{L}(B_1)_3=0$ yields
\begin{equation}\label{eq:condF1missed}
    k_1\cdot p_3=-\frac{m_2^2+M_3^2}{2}\,.
\end{equation}
Imposing \eqref{eq:condF1missed} on  $\det \tilde{\mathcal{G}}_1=0$ from \eqref{eq:Gtilde} with $\mathscr{S}=\{\mathscr{L}(B_1)_3\}$ gives (instead of \eqref{eq:detToSolve}):
\begin{equation}\label{eq:hyperint}
\hspace{-0.3cm}
\small
\begin{vmatrix}
 s & \frac{m_2^2-m_1^2+s}{2} & \frac{M_4^2-M_3^2-s}{2} \\
 \frac{m_2^2-m_1^2+s}{2} & m_2^2 & -\frac{m_2^2+M_3^2}{2} \\
 \frac{M_4^2-M_3^2-s}{2}  & -\frac{m_2^2+M_3^2}{2} & M_3^2 
    \end{vmatrix}=0\,,
\end{equation}
which agrees perfectly with $\texttt{D}[3]$. Such singularities were described as ``mixed-type" in the old literature \cite{osti_4773607,Eden:1966dnq} and their detection requires a blow-up in $\alpha$-parameter space \cite{Berghoff:2022mqu}. Here we see that our recursive method obtains all the parachute leading singularities without using blow-ups.

\redd{
In this simple yet non‑trivial example, we did not need the derivatives in \eqref{eq:LAB}, since imposing the constraint $\{\mathscr{S}_i=0\}$ eliminated all integration variables in \eqref{eq:example1} (i.e.,\ $\dim\,X_\mathscr{S}=0$ in every case). While one might invoke the derivative constraint for $\mathscr{S}=\emptyset$, in this example that choice produces no new leading singularity—it only introduces \texttt{D[9]} in \cite{data1}, which is non‑leading—and is therefore omitted from the discussion. In App.\,\ref{app:example}, we discuss a massive pentabox diagram for which derivatives cannot be avoided.
}
\vspace{0.1cm}
\myparagraph{More parachutes.}
The method similarly determines the (e.g., leading) singularities of $L$-loop diagrams where one adds propagators joining the same vertices as $\mathscr{D}_3$ and $\mathscr{D}_4$. The only change occurs in \eqref{eq:condF1} and simply requires replacing this condition by
\begin{equation}\label{eq:condF1L-loop}
    \hspace{-0.3cm}
 k_1\cdot p_3{=}\frac{1}{2}\big[( m_3\pm \ldots \pm m_{3+L})^2-m_2^2-M_3^2\big]\,,
\end{equation}
which is derived from the known singularities of the $L$-loop banana graph \cite[Prop.~2]{Mizera:2021icv} (see also \cite{Klemm:2019dbm}).
\section{\label{sec:examples}Checks and new predictions}
In Tab.~\ref{fig:checked} and Tab.~\ref{fig:predictions} we enumerate non-trivial diagrams whose leading singularities we explicitly calculated.
We stress-tested our method against 
and current state-of-the-art tools \texttt{PLD.jl} \cite{Mizera:2021icv,Fevola:2023fzn,Fevola:2023kaw} and \texttt{HyperInt} \cite{Brown:2009ta,Panzer:2014caa}, 
which utilize modern algebro-geometric and advanced numerical polynomial solving methods.
We recover every known singularity and also make new predictions.

In Tab.~\ref{fig:checked} we display diagrams for which we reproduced all known (leading) singularities predicted by these tools. For the diagrams where comparison with \texttt{HyperInt} was possible, we match exactly. For the diagrams where we could only compare with \texttt{PLD.jl}, we predict an equal or larger list of singularities, given explicitly in the ancillary file.
The additional singularities are of the mixed-type described above in the parachute example. \redd{For the examples in Tab.\,\ref{fig:checked} we found it sufficient to consider sets $\mathscr{S}$ that contain at most one element from each of $\mathscr{L}(A)$ or $\mathscr{L}(B)$; otherwise, no new singularities were found.}

For the diagrams in Tab.~\ref{fig:predictions}, no previous results were available to our knowledge. 
The full list of singularities, with the new ones highlighted, can be found in 
an ancillary Mathematica notebook available on GitHub \href{https://github.com/StrangeQuark007/recursive_landau}{\faGithub} \cite{repo}.
Although we do not have a rigorous proof that our method always finds the complete list of leading singularities of a given diagram, we take the perfect agreement with \texttt{HyperInt} on similar diagrams with fewer scales as suggestive evidence that our predicted lists are complete.

\begin{table*}[ht]
\centering
\begin{tabular}{ccc}
  \adjustbox{valign=c}{\includegraphics[scale=1]{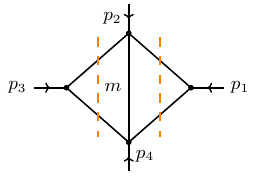}} &
  \adjustbox{valign=c}{\includegraphics[scale=1]{
  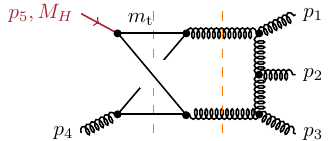
  }} &
  \adjustbox{valign=c}{\includegraphics[scale=1]{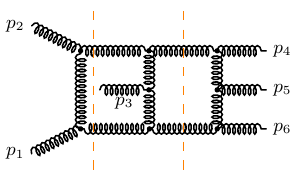}} \\
  \emph{Massive acnode} ($p_i^2=M^2$)
    & \emph{H+J pentabox \#1}  ($p_5^2=M_H^2\,, p_{i<5}^2=0$) 
& \emph{Nonplanar d-pentagon \# 1} ($p_i^2=0$)
\\
  \adjustbox{valign=c}{\includegraphics[scale=1]{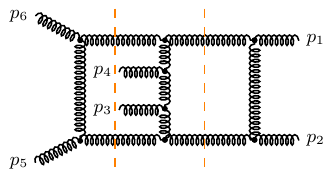}} &
  \adjustbox{valign=c}{\includegraphics[scale=1]{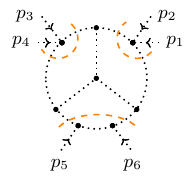}} &
  \adjustbox{valign=c}{\includegraphics[scale=1]{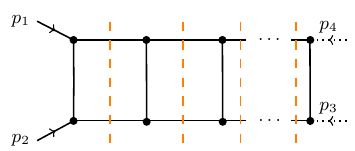}} 
  \\
 \emph{Nonplanar d-pentagon \# 2} ($p_i^2=0$)
& \emph{Mercedes diagram} ($p_i^2=0$)
& \emph{Massive ladder} ($p_{1,2}^2=m$\,, $p_{3,4}^2=0$)
\\
\end{tabular}
  \caption{
  \red{Non-trivial diagrams for which we recover all leading singularities in the literature and in some cases predict new ones.}
  The dotted orange lines show the two-particle cuts we used in the recursion. The kinematic conventions and results can be found in the ancillary file. \red{Although} we draw the diagrams to be evocative of Standard Model processes, the only important aspect is the mass of each edge (dotted and spiral lines represent massless particles, while the solid lines are massive).}
    \label{fig:checked}
\end{table*}
\begin{table*}[ht]
\centering
\begin{tabular}{ccc}
  \adjustbox{valign=c}{\includegraphics[scale=1]{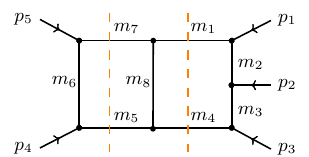}} &
  \adjustbox{valign=c}{\includegraphics[scale=1]{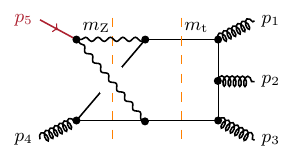}} &
  \adjustbox{valign=c}{\includegraphics[scale=1]{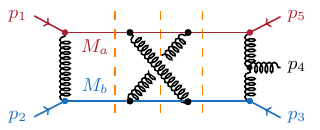}} \\
  \emph{Generic kinematic pentabox} ($p_i^2=M_i^2$)
& \emph{H+J pentabox \#2} ($p_5^2=M_H^2\,, p_{i<5}^2=0$) 
& \emph{QCD process} ($p_{1,5}^2=M_a^2\,, p_{2,3}^2=M_b^2$, $p_4^2=0$) 
\\
  \adjustbox{valign=c}{\includegraphics[scale=1]{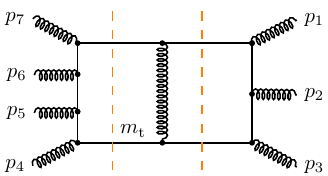}} &
  \adjustbox{valign=c}{\includegraphics[scale=1]{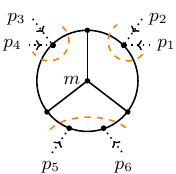}} &
  \adjustbox{valign=c}{\includegraphics[scale=1]{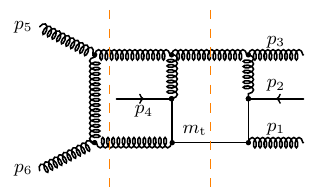}} \\
  \emph{Massive hexapentagon}  ($p_i^2=0$)
& \emph{Massive Mercedes diagram}  ($p_i^2=0$)
& \emph{Hexabox}  ($p_{2,4}^2=m_t^2$, $p_{1,3,5,6}^2=0$)
\\
& \adjustbox{valign=c}{\includegraphics[scale=1]{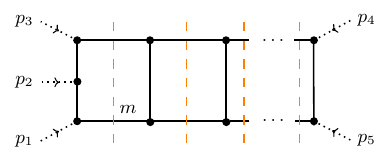}} & \\ 
& \emph{Massive pentaladder}  ($p_i^2=0$)
\end{tabular}
  \caption{
Diagrams for which we have predicted the leading singularities and for which no previous results were available. 
(Dotted and spiral lines indicate massless particles, while the solid and wavy lines are massive.)}
    \label{fig:predictions}
\end{table*}
\section{\label{sec:conclusion}Conclusion}
In this letter, we explored how unitarity can be used to extract recursively the Landau singularities of Feynman graphs. Our master formula, \eqref{eq:LAB}, relates the Landau singularities of graph $AB$ to those of subgraphs $A$ and $B$, which are disconnected by a two-particle cut. The recursive nature of our method, together with its direct application in momentum space, makes it remarkably practical and efficient for a vast class of examples.

We showed in detail how the two-loop parachute leading singularity can be obtained recursively from the bubble subgraph's leading singularity. We also illustrated how the method captures other types of singularities (some known as \emph{second-type} \red{in the Landau analysis literature \cite{Eden:1966dnq,osti_4773607}}) by performing the same recursion using corresponding singularities of the subgraphs, bypassing known subtleties in conventional Landau analysis regarding blow-ups in Schwinger parameter space \cite{Berghoff:2022mqu, Fevola:2023fzn, Fevola:2023kaw}.

We stress-tested our method on two-loop diagrams for which state-of-the art tools could predict (sometimes incomplete) lists of singularities
(see Tab.~\ref{fig:checked}),
as well as diagrams
(see Tab.~\ref{fig:predictions}) that \red{none of the current state-of-the-art} 
tools could solve in useful time. 
Many of these examples are relevant to Standard Model processes involving massive quarks and electroweak bosons. All of our calculations were performed analytically on a conventional laptop using built-in \textsc{Mathematica} functions. The results are available in the repository: 
\href{https://github.com/StrangeQuark007/recursive_landau/tree/main}{\faGithub} \cite{repo}.

\begin{figure}
    \centering
    \resizebox{0.45\textwidth}{!}{
    \includegraphics[scale=1]{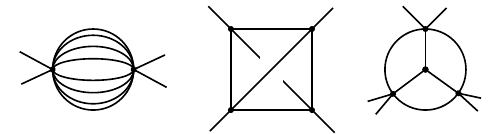}
    }
    \caption{Examples of loop diagrams with \emph{no} two-particle cuts splitting the graph in two disconnected subgraphs.}
    \label{ex:nonExample}
\end{figure}


In the examples considered here, the system of equations \eqref{eq:LAB} resulted in relatively low-degree polynomial constraints. This might change for more complicated Feynman diagrams, where more refined algebraic geometry methods could be used to compute the discriminant of the system more efficiently. Currently, the complexity and number of polynomial equations \red{appear} manageable thanks to the recursive nature of the method, the absence of variables other than kinematic invariants, \red{and possibly the two-particle-reducible nature of the considered diagrams.
An important question for the future is whether more complicated diagrams will require blow-ups in momentum space.}

Plausible generalizations of \eqref{eq:LAB} to three-particle cuts could involve replacing $\det \mathcal{G}$ either with a larger Gram determinant that includes the additional intermediate momentum, or with three independent Gram constraints $\det \mathcal{G}(k_i \cdot P_{AB}') = 0$ for each independent cut propagator momentum $k_i$ \cite{Gribov:1962ft, kim}. In practice, this will only be \red{necessary at three or higher loops}
because the current recursion terminates on two-particle-irreducible diagrams, which can be readily classified; see Fig.~\ref{ex:nonExample}. 


In some situations one might want to know whether a given singularity can appear on the principal sheet, \red{in particular for bootstrap applications or dispersive approaches}.  Since our method in principle constructs momenta (and potentially corresponding Schwinger parameters) that solve the Landau equations, we anticipate that such more refined questions \red{may} also be answered recursively.

We anticipate that this work will lead to computational advances in analytic studies of Feynman integrals \red{where knowledge of Landau singularities is often a pre-requisite, as mentioned in the introduction. 
We also envision potential applications to other areas of particle physics where Landau analysis is relevant such as cosmological correlators \cite{Goodhew:2021oqg,Salcedo:2022aal,Lee:2023kno,Donath:2024utn,Ema:2024hkj}, 
or inclusive/out-of-time-ordered amplitudes \cite{Caron-Huot:2023vxl}.}

\redd{\emph{Note added.} After the initial submission of this letter, two publicly available packages have appeared \cite{Frellesvig:2024ymq,Correia:2025yao} which generate the $\mathcal{G}$ polynomials entering \eqref{eq:LAB} via the loop‐by‐loop Baikov representation. Building on the work presented in this letter, the package \cite{Correia:2025yao} automatically performs the singularity analysis, including diagrams that lie beyond the two‐particle–reducible class.}


\acknowledgments We thank Lance Dixon, Giulia Isabella, Hofie Hannesdottir, Andrew McLeod, Sebastian Mizera, and Alexander Zhiboedov for useful discussions and comments on the draft. We also thank Sebastian Mizera for answering our questions on \texttt{PLD.jl}.
The work of S.C.H. is supported by the National Science and Engineering Council of Canada (NSERC), the Canada Research Chair program, reference number CRC-2022-00421.
Both S.C.H. and M.C. are supported by the Simons Collaboration on the Nonperturbative Bootstrap.  M.G.’s work is supported by the National Science and Engineering Council of Canada (NSERC) and the Canada Research Chair program, reference CRC-2022-00421.

\appendix
\section{Massive pentabox example}\label{app:example}
The purpose of this section is to demonstrate once more how the method works in practice through a detailed analysis of a known, yet non-trivial, example. We will consider the pentabox with two loops of equal mass $m$
\begin{equation}
    \adjustbox{valign=c,scale={1}{1}}{
    \includegraphics[scale=1]{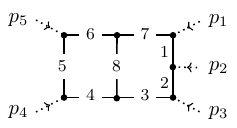}
    }
\end{equation}
where $p_i^2=0$ for all $i$. We use the following set of propagators:
\begin{align}
    \hspace{-0.4cm}\begin{aligned}
        &\mathscr{D}_1=(k_1{-}p_{23})^2{-}m^2\,, &&\mathscr{D}_2=(k_1{-}p_{3})^2{-}m^2\,,
        \\ 
        &\mathscr{D}_3=k_1^2{-}m^2\,, &&\mathscr{D}_4=k_2^2{-}m^2\,,
        \\
        &\mathscr{D}_5=(k_2{-}p_4)^2{-}m^2\,, &&\mathscr{D}_6=(k_2{+}p_{123})^2{-}m^2\,,
        \\  
        &\mathscr{D}_7=(k_1{-}p_{123})^2{-}m^2\,, &&\mathscr{D}_8=(k_2{+}k_1)^2{-}m^2\,.
    \end{aligned}
\end{align}

The recursion begins with the box (contained in the inner \textcolor{Maroon}{red} rectangle) and ends with the full graph (contained in the outer \textcolor{RoyalBlue}{blue} rectangle)
\begin{equation}
    \adjustbox{valign=c,scale={1}{1}}{
    \includegraphics[scale=1]{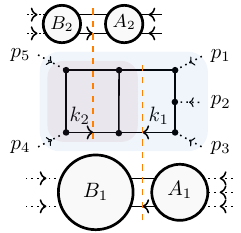}
    }
\end{equation}
(Note that starting with the pentagon yields the same set of singularities in the end, although \eqref{eq:LAB} is harder to solve since it involves one derivative condition.)

For the box, we have the Gram matrix
\begin{equation}\label{eq:G1}
    \mathcal{G}_2=\Big[q_i^{(2)}\cdot q_j^{(2)}\Big] \quad :~q_\bullet^{(2)}\in\{p_4,k_2,k_1,p_{123}-k_1\}\,.
\end{equation}
We can eliminate the internal dot products
\begin{equation}
        k_2\cdot\{p_4,k_2,k_1,p_{123}-k_1\}\,,
\end{equation}
by first localizing on the two-particle cut $\mathscr{D}_4 {=} \mathscr{D}_6 {=} 0$ and then localizing on the leading singular loci $\mathscr{L}(A_2)_1{=}0$ and $\mathscr{L}(B_2)_1{=}0$ of ${A_2}$ and ${B_2}$, given by $\mathscr{D}_5 {=} \mathscr{D}_8 {=} 0$. The result is the restricted Gram determinant $\det \tilde{\mathcal{G}}_2$ from \eqref{eq:Gtilde} with $\mathscr{S}=\{\mathscr{L}(A_2)_1,\mathscr{L}(B_2)_1\}$, which explicitly reads
\begin{equation}\label{eq:condF1pb}
   \hspace{-0.35cm} \footnotesize \begin{vmatrix}
        0 & 0 & k_1 {\cdot} p_4 &p_4 {\cdot} p_{123} {-} k_1 {\cdot} p_4 \\
 0 & m^2 & {-}\frac{k_1^2}{2} & \frac{k_1^2 {-} p_{123}^2}{2} \\
 k_1{\cdot} p_4 & {-}\frac{k_1^2}{2} & k_1^2 & k_1 \cdot p_{123} {-} k_1^2\\
p_4 {\cdot} p_{123} {-} k_1 {\cdot} p_4 & \frac{k_1^2{-} p_{123}^2}{2} & k_1 {\cdot} p_{123} {-} k_1^2 &  k_1^2 {+} p_{123}^2{-}2 k_1 {\cdot} p_{123}
    \end{vmatrix}\,.
\end{equation}
Setting \eqref{eq:condF1pb} to zero gives a constraint $(\mathscr{L}(B_1)_1{=}0)$ to impose in the next step of the recursion.

For the full graph, we have the Gram matrix
\begin{equation}
    \mathcal{G}_1=\Big[q_i^{(1)}\cdot q_j^{(1)}\Big] \quad :~q_\bullet^{(1)}\in\{p_3,p_{23},p_{123},k_1,p_4\}\,.
\end{equation}
To eliminate all but one internal dot product
\begin{equation}
    k_1\cdot\{p_3,p_{23},p_{123},k_1,p_4\}\,,
\end{equation}
we first localize on the two-particle cut $\mathscr{D}_3 {=} \mathscr{D}_7 {=} 0$ and then on the leading singular loci $\mathscr{L}(A_1)_1{=}0$ and $\mathscr{L}(B_1)_1{=}0$ of ${A_1}$ and ${B_1}$. $\mathscr{L}(A_1)_1{=}0$ is given by $\mathscr{D}_1 {=} \mathscr{D}_2 {=}0$, which, together with the two-particle cut, fix $k_1^2$, $k_1\cdot p_{123}$, $k_1\cdot p_3$ and $k_1\cdot p_{23}$. To fix $k_1\cdot p_4$, we solve $\mathscr{L}(B_1)_1=\eqref{eq:condF1pb}=0$. This yields the restricted Gram determinant $\det \tilde{\mathcal{G}}_1$ from \eqref{eq:Gtilde} with $\mathscr{S}=\{\mathscr{L}(A_1)_1,\mathscr{L}(B_1)_1\}$
\begin{equation}\label{eq:finalDetptb}
 \hspace{-0.4cm}  
 \begin{vmatrix}
0 & \frac{s_{23}}{2} & \frac{s_{45}{-}s_{12}}{2} & 0 & \frac{s_{34}}{2} \\
 \frac{s_{23}}{2} & s_{23} & \frac{s_{23}{+}s_{45}}{2} & \frac{s_{23}}{2}  & \frac{s_{15}{-}s_{23}}{2} \\
 \frac{s_{45}{-}s_{12}}{2} & \frac{s_{23}{+}s_{45}}{2} & s_{45} & \frac{s_{45}}{2} & -\frac{s_{45}}{2} \\
 0 & \frac{s_{23}}{2} & \frac{s_{45}}{2} & m^2 &\frac{m^2 s_{45}^2{\pm}2 \sigma}{\eta} \\
 \frac{s_{34}}{2} & \frac{s_{15}{-}s_{23}}{2} & -\frac{s_{45}}{2} & \frac{m^2 s_{45}^2{\pm} 2 \sigma}{\eta} & 0 
    \end{vmatrix}\,,
\end{equation}
where $\sigma=\sqrt{m^2 s_{45}^3 (s_{45}-3m^2)}$ and $\eta=2 s_{45}(s_{45}{-}4 m^2)$. Because the result is independent of $k_1$, there is no derivative condition to impose and $\mathscr{L}(A_1B_1)_1{=}0$ is given by setting $\eqref{eq:finalDetptb}$ equal to zero. The result agrees with $\texttt{D[57]}$ in the database \cite{data2}, which is one of the leading singularity components flagged by \texttt{PLD.jl}. The other one, $\texttt{D[62]}$, arises from the kinematic normalization $C \propto (\det[p_i \cdot p_j]_{i,j=1}^4)^{\frac{5-\D}{2}}$ in \eqref{eq:cut}.

\bibliography{references}
\end{document}